\documentstyle[aps,multicol,epsfig]{revtex}
\begin{document}
\draft
\title{Similarity of slow stripe fluctuations between Sr-doped cuprates and
oxygen-doped nickelates.}

\author{I.M. Abu-Shiekah, O.O. Bernal\cite{OOB}, A.A.  Menovsky\cite{AM}, and
H.B. Brom}

\address{Kamerlingh Onnes Laboratory, Leiden University,\\
         P.O.Box 9506, 2300 RA Leiden, The Netherlands}

\author{J. Zaanen}

\address{Instituut Lorentz for Theoretical Physics, Leiden University,\\
P.O. Box 9504, 2300 RA Leiden, The Netherlands}

\date{June 14, 1999}

\maketitle

\begin{abstract}
Stripe fluctuations in La$_2$NiO$_{4.17}$ have been studied by $^{139}$La NMR
using the field and temperature dependence of the linewidth and relaxation
rates.  In the formation process of the stripes the NMR line intensity is
maximal below 230~K, starts to diminish around 140~K, disappears around 50~K
and recovers at 4~K.  These results are shown to be consistent with, but
completely complementary to neutron measurements, and to be generic for
oxygen doped nickelates and underdoped cuprates.
\end{abstract}


\begin{multicols}{2}
\settowidth{\columnwidth}{aaaaaaaaaaaaaaaaaaaaaaaaaaaaaaaaaaaaaaaaaaaaaaaaa}

Evidence is accumulating that the electron systems in doped Mott-Hubbard
insulators exhibit quite complex ordering phenomena\cite{Emery93}.  In two
dimensional (2D) systems this takes the form of stripe phases, where the
excess charges bind to antiphase boundaries in the N\'eel
state\cite{Zaanen89}.  It was recently demonstrated by
Hunt {\sl et al.}\cite{Hunt99} that in a large temperature regime where the
stripe order appears to be complete according to diffraction experiments, the
stripe system is still slowly fluctuating. This follows from NMR experiments,
showing both motional narrowing at higher temperatures and a wipe-out of the
NMR signal upon cooling down, caused by the characteristic fluctuation
frequency becoming of the order of the NQR linewidth/splitting of a few MHz.
Here we will demonstrate that these fluctuations are not unique to cuprate
stripes, and thereby unrelated to intricacies associated with the proximity
of the superconducting state. We present a NMR study of the  stripe phase in
oxygen doped La$_2$NiO$_4$. It seems well established that the excess oxygen
enters as an interstitial that shows a tendency to order three dimenionally,
creating a larger unit
cell\cite{Jorgensen89,Rodriguez91,Mehta94,Tranquada97}. It should therefore
be regarded as a rather clean system compared to the Sr-doped
nickelates\cite{ABZ99}.  We find that this `clean' nickelate system exhibits
a fluctuation behavior which closely parallels the fluctuations of the
cuprate stripes: although scattering experiments in
La$_2$NiO$_{4.13}$\cite{Tranquada97} show charge- and spin freezing at
$T_{\rm CO}$=220~K and $T_{\rm SO}$=110~K, our NMR experiments in
La$_2$NiO$_{4.17}$ indicate that the stripes only become static at a
temperature of 2~K.  Interestingly, these slow fluctuations seem absent in
the 'dirty' Sr-doped nickelate
La$_{5/3}$Sr$_{1/3}$NiO$_4$\cite{Yoshinari99} (with the same 1/3 hole doping
content as our oxygen doped nickelate), where $\mu$SR\cite{Jestadt99}
measurements reveal the onset of static spin order at the same  temperature
(200~K) as the (quasi)elastic peaks develop in the neutron
scattering\cite{Lee97}.  These observations suggest that the slow stripe
fluctuations, characteristic for the cuprates and oxygen doped nickelates,
are in first instance unrelated to quenched disorder; at the same time, that
disorder is excessively effective in pinning these slow, intrinsic
fluctuations in the insulator but not in the (super)conductor.

Below we analyze the field and temperature dependence  of the $^{139}$La
linewidth and relaxation rates for
La$_2$NiO$_{4+\delta}$ with $\delta= 0.17$. $^{139}$La has a
nuclear spin $I=7/2$, which makes NMR sensitive to both charge and spin, and
allows the study of charge and spin order and also the dynamics at time
scales longer than $10^{-7}\, {\rm s^{-1}}$.  The measurements were performed
on two single crystals from different batches that were prepared under
atmospheric condition in a mirror oven at 1100~K\cite{Bernal97}.  Slices from
both samples were analyzed by microprobe techniques; on the average the
oxygen contents were found to be the same.  Samples for the measurements were
cut from those parts that had a homogeneous oxygen content
and had a typical weight of 50 mg.  Thermogravity (TGA) analysis
of the oxygen concentration gave $\delta$ = 0.17. The same value follows from
the volume of the unit cell as determined by X-ray\cite{Rice93}.  Samples of
batch 1 show a peak in the susceptibility ($\chi$) around 110~K, which is
also seen in the 2/15-doped compounds\cite{Tranquada97} and is associated
with oxygen order\cite{Poirot98}.  The absence of this peak in the
other sample (batch 2) points to local differences in the two samples.  Apart
from the peak, the susceptibility of both samples show the same Curie like
dependence on $T$ down to 80~K.  Below that temperature down to 4~K $\chi$
increases smoothly, resembling (but not identical to) Curie-Weiss behavior
with an antiferromagnetic exchange interaction. The $\chi$-data above
100~K give a Ni moment of about one $\mu_B$\cite{Bernal97,Poirot98}. The NMR
results of the two batches are qualitatively identical with small differences
in the freezing (see below) temperatures.  The results presented relate to
the samples of batch 2.

\begin{figure}[htb]
\begin{center}
\leavevmode
\epsfig{figure=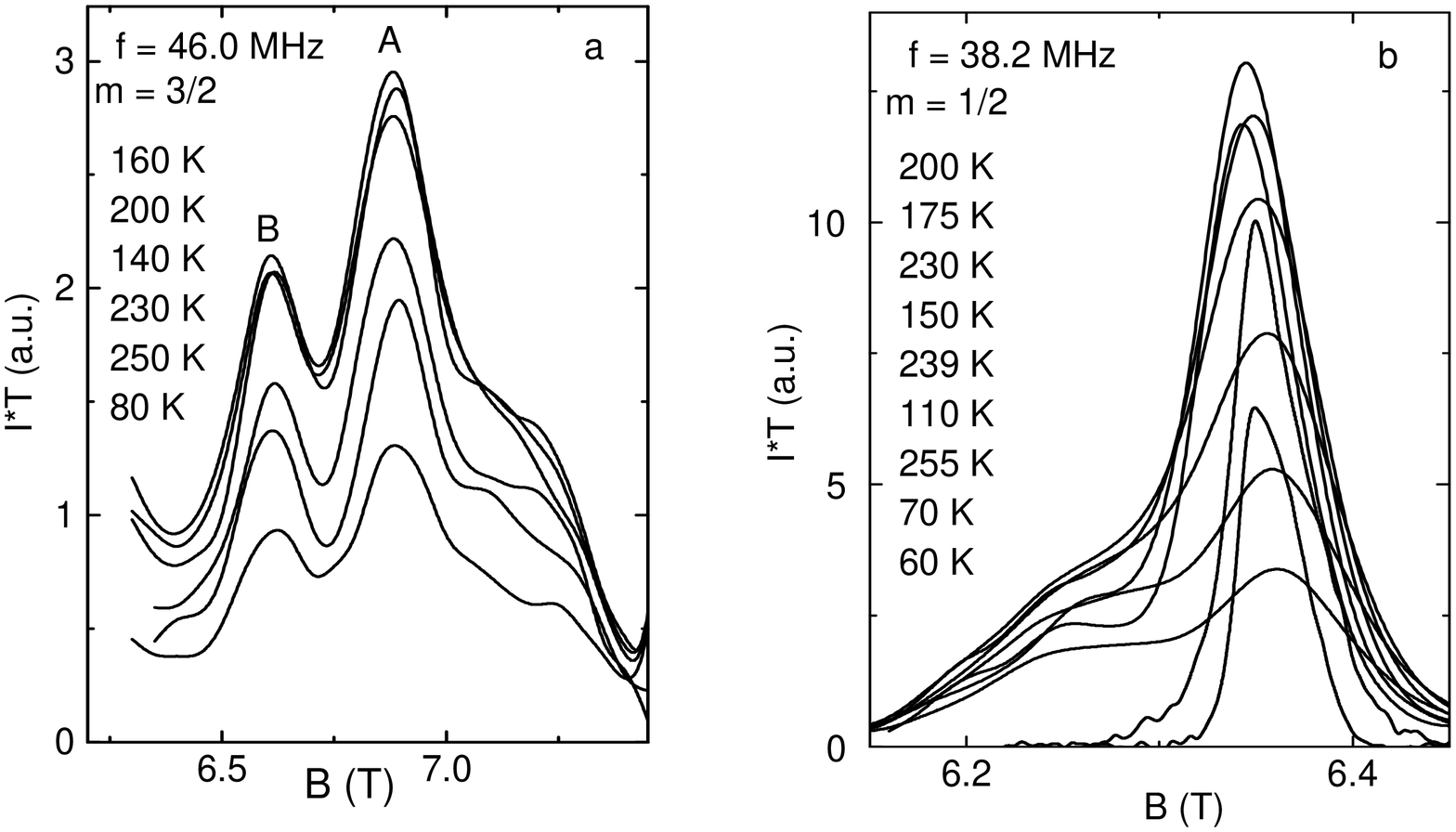,height=8.5cm,angle=-90}
\end{center}
\caption{$^{139}$La NMR field-profiles for $m=3/2$ (46.0 MHz) (a)
and $m=1/2$ (38.2 MHz) (b) and as a function of $T$. Temperatures
are listed according the maxima of the lines, e.g. in (a) the
highest maximum being for 160~K, the lowest for 80~K.  For $m=1/2$ below
230~K a second line (B) comes up on the low field side of
line A; for $m=3/2$ the two lines visible from the start (below 250~K).}
\label{f2}
\end{figure}

\begin{figure}[htb]
\begin{center}
\leavevmode
\epsfig{figure=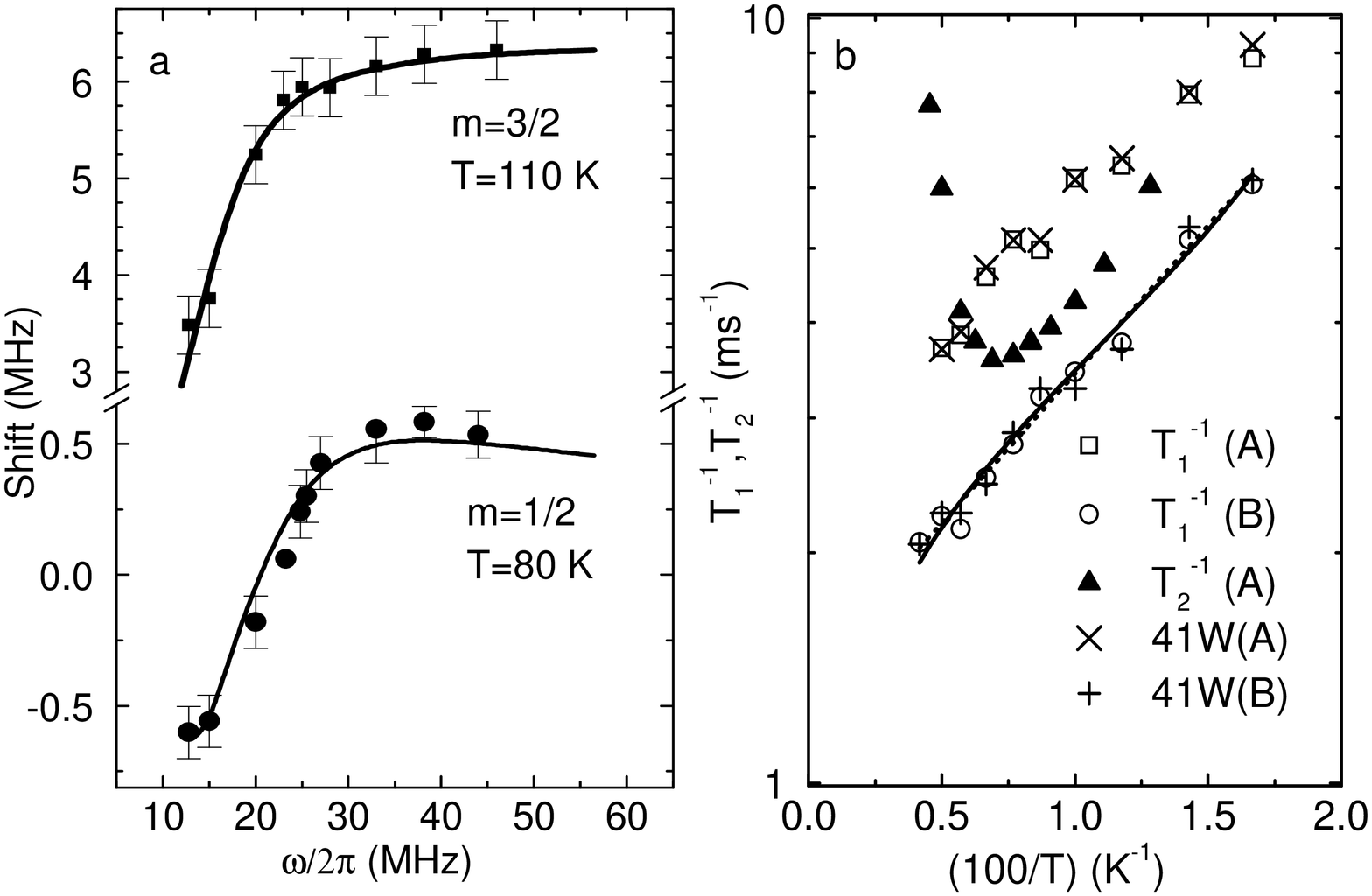,height=8.5cm,angle=-90}
\end{center}
\caption{(a) The lineposition of line B as function of resonance frequency
(and hence resonance field) for $B \parallel c$-axis.
The position of line A is field independent. Drawn line is a fit
with only quadrupolar interactions, which shows that the
lineposition and linewidths are due to charges.  (b) $T$-dependence of
$T_1^{-1}$(A), $T_1^{-1}$(B) and $T_2^{-1}$(A) $vs$ $T$.  The difference in
$T_1^{-1}$(B) and $T_2^{-1}$(A) is about a factor 2.  Drawn and dotted lines
are resp. fits to the spin freezing and activated model discussed in the
text.} \label{f4}
\end{figure}

\begin{figure}[htb]
\begin{center}
\leavevmode
\epsfig{figure=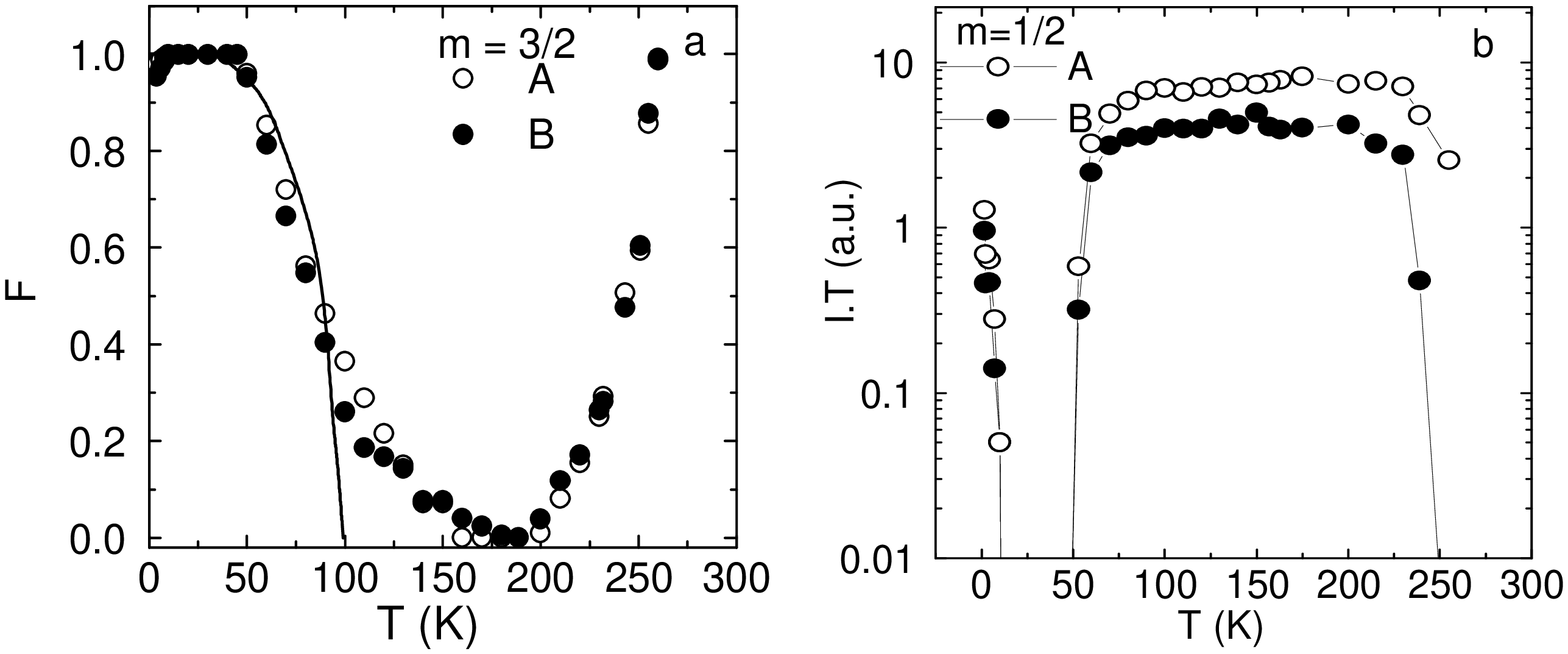,height=8.5cm,angle=-90}
\end{center}
\caption{(a) The $T$ dependence of $F$ for $m=3/2$.  The
intensities for lines A and B are normalized. Drawn line is a fit
to the BCS gap function. (b) Logarithmic plot of the intensities ($I$) times
$T$ for $m=1/2$. The maximal value of $I \cdot T$ of
line A is normalized to 1.}
\label{f5}
\end{figure}

\begin{figure}[htb]
\begin{center}
\leavevmode
\epsfig{figure=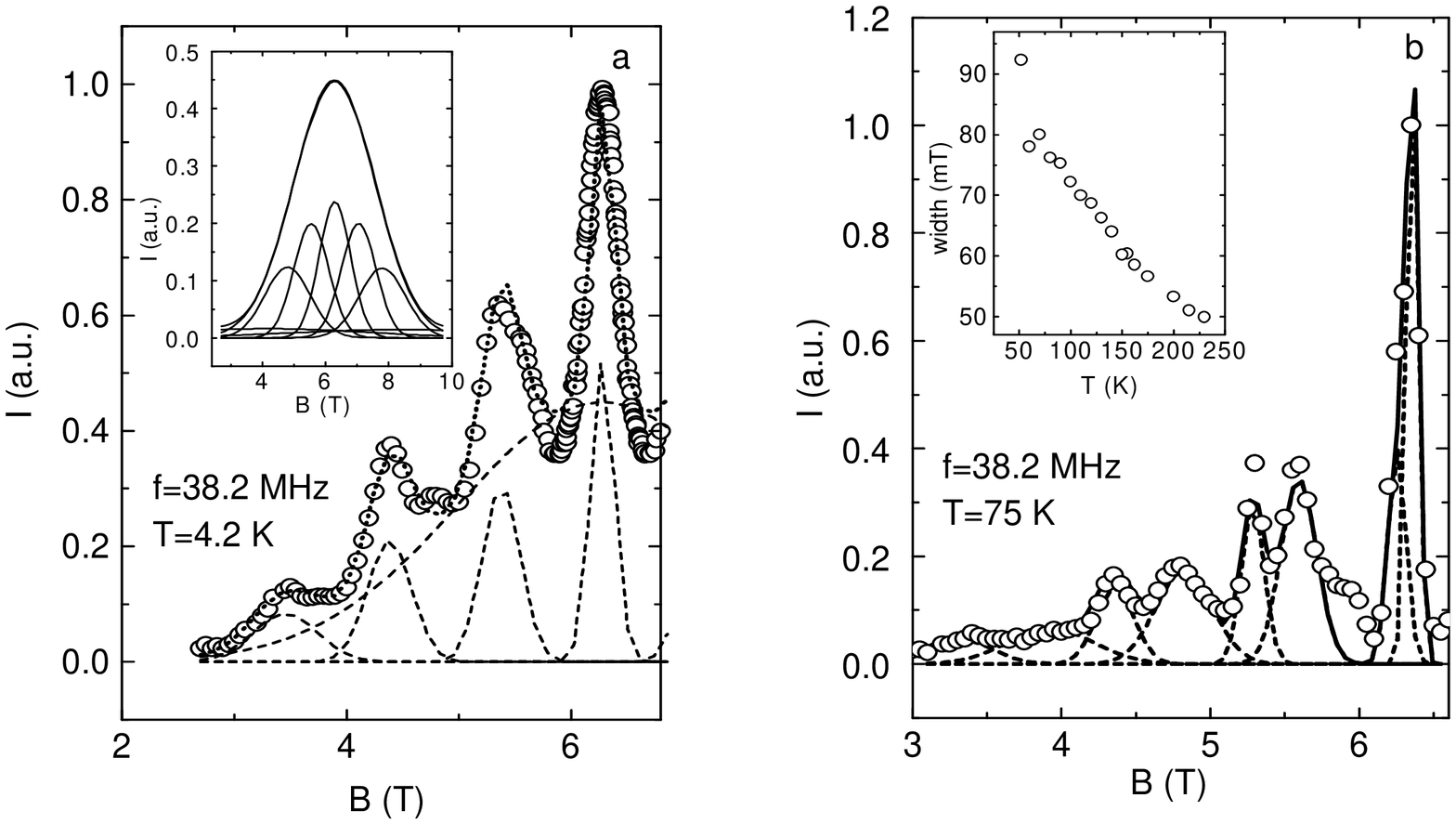,height=8.5cm,angle=-90}
\end{center}
\caption{Comparison of full sweeps at 4.2~K (a) and 75~K (b)
shows the extra line broadening at 4.2~K. The inset in (a) shows
the decomposition for the A line at 4.2 K; the inset in (b) gives
the $T$-dependence of the width of the A line for $m=1/2$. Down to $T=60$~K
the B-linewidth remains about 0.1 T (.6 MHz).}
\label{f3}
\end{figure}

Line profiles as function of temperature were determined by
frequency sweeps at fixed field (9.4~T and 4.7~T) or field sweeps
at fixed frequency. Both methods give the same results. As
detuning is not needed, the latter data have a better accuracy and are
analysed in the following.  Below 250~K, the $-3/2 \leftrightarrow -1/2$
($m=3/2$) transition becomes visible and is seen to be split into two lines,
A and B, Fig.\ref{f2}. In the same figure we show the results for the $-1/2
\leftrightarrow +1/2$ ($m=1/2$) transition, which develops a second component
at about 230~K\cite{CM98m}. As we will further clarify, these lines have the
same origin and correspond with La sites which are inequivalent because of
different electrical field gradients\cite{CM98}.  We are facing the ambiguity
that these differences can originate either in the distribution of the excess
oxygen, or that it might reflect the inhomogeneous charge distribution
associated with the stripes in the NiO$_2$ planes: in both cases a similar
pattern is expected\cite{ABZ99}.
At present we are further investigating these matters; important
for the present context is that both lines reveal a very similar temperature
dependence suggesting that both sites communicate with the same electronic
system as we will now demonstrate.
The field dependence of the two lines was measured in detail for
both transitions. Only the position of B depends on field as illustrated for
the $m=3/2$ and $m=1/2$ in Fig.\ref{f4}a.
The $T$-dependences of the spin-lattice relaxation rates ($T_1^{-1}$) are
shown in Fig.\ref{f4}b for the two $m=3/2$ lines,
together with the spin dephasing rate ($T_2^{-1}$) for the A line.
$T_1^{-1}$ is measured by a $\pi$--$t$--$\pi/2$--$\tau$--$\pi$ sequence and
analyzed with the multi-exponential expression of
Narath\cite{Narath67}.  The effective relaxation rates correspond
to 41$W$, the fundamental transition probability.  The magnetic character of
the relaxation mechanism below 200~K was checked by comparing the relaxation
rates for the $m=1/2$ and $m=3/2$ transitions.  When analyzed by the magnetic
expression they gave the same $W_M$.  $T_1^{-1}$(A) is about twice as fast as
$T_1^{-1}$(B).
The intensity ratio of B to A depends on the cooling history,
being larger when cooling proceeds slower.  In Figs.~\ref{f2}-\ref{f5} we
show the results obtained after slowly cooling down.  With decreasing
temperature the intensity of the resonance lines first increases, then
saturates around 150~K to disappear completely around 50~K.  Fig.\ref{f5}
gives the $T$ dependence of the normalized intensity (log.  scale) and of the
wipe-out fraction $F$ (the intensities are integrated over the linewidth,
corrected for the apparent $T_2$'s), defined as the ratio of the
experimentally observed and the Curie extrapolated value.
At lower temperatures the A and B lines resurrect again; at 4.2~K
the linewidths at 38.2 MHz (A $\sim 6$~MHz, B $\sim 2$~MHz) are
about 3 times larger than at 75~K, see fig.\ref{f3}.  Also in zero field line
B ($m=7/2$) is only broadened, not split, with a width which is comparable
(3~MHz) to that found in field for the same transition.  This width is an
order of magnitude higher than the calculated effect of the dipolar field.
There is only a faint $T$-dependence in the position of the two lines. Above
50~K the width of A is increased by 50\%, while that of line B is even less
$T$ dependent.

Let us now turn to the interpretation of the data.  Consistent with the
neutron and susceptibility measurements in
La$_2$NiO$_{4.13}$\cite{Tranquada97}, we find in La$_2$NiO$_{4.17}$ the
following temperature regimes:

(i) $T > 250$~K. Only a narrow $m=1/2$ NMR line is visible with strongly
reduced intensity.  All visible La-sites experience the
same (averaged) electrical field gradient. Oxygen motion explains
the activated relaxation rates of the $m=1/2$ lines (not shown here, see
ref.\cite{CM98m}) with an activation energy of $3\times 10^3$~K, and the
increase of $T_2^{-1}$(A), shown in fig.\ref{f4}.

(ii) 250~K$ >T > $230~K.  Due to oxygen/charge ordering the
$m=3/2$ transition comes up with two lines A and B with an intensity ratio
close to 2. Due to the small intensity this splitting is not yet visible in
the $m=1/2$ line.

(iii) 230~K$>T>$140~K. Below 230~K the same two inequivalent sites
(A and B), that can be distinguished in the $m=3/2$ transition, broaden the
$m=1/2$ line, Fig.~\ref{f2}.  The field dependence of
the line pattern is well explained by a quadrupolar interaction, drawn lines
in Fig.\ref{f4}\cite{ABZ99}. The saturation of the intensity ratio of two to
one for the unshifted to shifted line around 230~K is expected for site
ordered stripes\cite{CM98m,CM98}.

(iv) 140~K$>T>$50~K. Slow motion of spins or charges wipes out
most of the NMR intensity. The nuclei that can still be observed
show additional line broadening and a relaxation rate, which is reminescent
to the $T$-dependence of $T_1^{-1}$ in
La$_2$Cu$_{1-x}$Li$_x$O$_4$\cite{Suh98} or Sr doped
La$_2$CuO$_4$\cite{Chou93} above the spin freezing temperature. $T_1^{-1}(T)$
can be fitted with an activated process $T_1^{-1} \propto \exp (\Delta
E_a/T)$ with $\Delta E_a = 10^2$~K (dotted line in Fig.~\ref{f4}) or a power
law dependence $T_1^{-1} \propto [(T-T_f)/T_f]^{\alpha}$ with
$\alpha \sim -0.46$ (drawn line in Fig.~\ref{f4}) and a spin freezing
temperature $T_f$ of about 45~K.  As the $T$-dependence of the rates of A and
B spins look identical, the ratio of a factor 2 is due to the strength of the
hyperfine coupling to the electron spins.  When analyzed in the renormalized
classical limit\cite{Tye89}, the activation energy equals the exchange
constant $J$.  The magnetic correlation length $\xi/a = 0.5 \exp(J/T)$ above
$10^2$~K is of the order of the lattice constant $a$, and an order of
magnitude larger around 40~K. In any case the relaxation rates are not
related to the wipe-out process, see Figs.\ref{f5},\ref{f4}.  The same
behavior is observed in the cuprates and it suggests\cite{Hunt99} that the
bulk of the stripe system is subjected to fluctuations on the MHz scale (the
invisible fraction) while local events occur characterized by spin
fluctuations on a much shorter time scale. Finally, it appears that the
functional depenence of the wipe-out factor $F$ on $T$ is similar to that of
the $s$-wave BCS order parameter (see Fig.~\ref{f5})\cite{Hunt99}.

(v) 50~K$>T>$15~K. Wipe out of the NMR intensity is complete.
The whole La-nuclear system experience the fluctuations on MHz
time scales and the rare sites dealt with in the previous
paragraph have disappeared completely.

(vi) 15~K$>T$. After emerging again, the La NMR lines are
appreciably broadened with a width that is at least a few times larger for  A
than for B spins. This signals the onset of truly static order and it is
noticed that even at the lowest
temperatures where we measured (2~K) only 10\% percent of the intensity has
recovered. The line width reflects the static
disorder (inhomogeneous broadening) and one would be tempted to ascribe this
to spin glass behavior.  Interestingly, we find that
the width of $m=3/2$ does not depend on the field strength as expected for
disorder originating in the charge sector\cite{Kivelson98}.

When we compare our data with NMR data in underdoped La$_{2-
x}$Sr$_x$CuO$_4$\cite{Hunt99}, the similarities in the wipe
out process are striking. In all samples stripe structures are clearly seen
in the neutron data, but seem to be absent in the
NMR data.  This apparent contradiction is solved if the wipe-out effect seen
in the NMR data is associated with slow dynamics
in the stripe system.  Almost all cuprates, studied in ref.\cite{Hunt99}, are
superconductors - the only non-superconducting sample with 0.04Sr content
behaving anomalously as it shows no stripe fluctuations.
In the cuprates at low doping, where the superconducting fraction seems to be
very low,  the wipe out fraction starts to grow
typically below 50~K and equals $\sim 1$ around 20~K. Near optimal doping
(0.15Sr content) $F$ is $\leq 0.3$\cite{Hunt99}.  In the underdoped cuprates
static magnetic hyperfine fields changes the NQR frequencies completely below
10~K\cite{Hunt99}, the same ordering temperature as seen by
$\mu$SR\cite{Niedermayer98}.  Here the comparison between cuprates and
nickelates is elucidating.  The nickelates are very poor
conductors\cite{odier99} and show no superconductivity.  As mentioned above,
in neutron scattering on the nickelates with oxygen doping of
$0.133$\cite{Tranquada97} charge order is observed around 210~K while spins
order around 110~K: we find that static order only occurs at temperatures
less than 10 K.  The wipe out fraction of one is in agreement with the trend
in the cuprate data. The onset of the wipe-out process around 140~K makes the
combined action of spin and charge motion in the already formed stripes
structure a plausible explanation for its origin. More specifically, the
general pattern seems susceptible to an explanation in terms of dislocation
melting. As is well understood in the context of two dimensional
melting\cite{Kosterlitz73}, dislocation unbinding leads to an overall fluid
character of the system, and such a transition can occur also when the
density of topological defects is low, i.e. in a regime where the time scales
are large.  However, we also found the fast spin relaxations in the fraction
of the system which is still visible in NMR when the wipe out is nearly
complete. We are tempted to ascribe those to the dislocations: we envisage
that in the core region of the dislocations the spin system is strongly
frustrated.

We end by mentioning the puzzling aspect related to the role of quenched
disorder in the Sr doped nickelates, compared to oxygen doped systems we have
been studying. Especially for hole doping of 1/3, the difference between
oxygen (O4.17) and Sr doped (Sr0.33) nickelates is manifest\cite{Jestadt99}:
in the neutron scattering spin ordering in O4.15 is seen around 110~K (the
wipe out process in our samples starts around 140~K), while in Sr-doped
samples this temperature is around 200~K (both in $\mu$SR, NMR and NS).
The differences go further. No strong wipe-out features are
present for Sr0.33\cite{Yoshinari99}, and the intensity ratio of
the two NMR lines differs from ours. We believe these effects to
be linked to the effect of the "quenched" Sr-dopant in this insulator, and we
find an analogy with what happens in highly two-dimensional type II
superconductors.  As is well known in that field, pinning centers can
suppress the formation of a flux liquid, i.e enhance the vortex melting
temperature appreciably above the Kosterlitz-Thouless melting temperature.

In summary, our NMR study of stripes in oxygen doped La$_2$NiO$_4$ shows a
striking similarity with the behavior found previously in cuprate
superconductors. The stripe system is  apparently a slowly fluctuating
strongly correlated fluid over an extended range of temperatures. Since this
behavior also occurs in the insulating nickelate, these fluctuations are
uncorrelated to the proximity with the superconducting state, while it also
appears as unlikely that it is completely driven by quenched disorder. We
suspect that our findings are related to the peculiar character of the stripe
phase itself.

We gratefully acknowledge fruitful discussions with S. Mukhin about
theoretical aspects, with P.C. Hammel about the NMR results in
La$_{5/3}$Sr$_{1/3}$NiO$_4$, and D.E. MacLaughlin about NMR in spin-glass
systems.  One batch of the single crystals was prepared
by Y.M.  Mukovskii at the Steel and Alloys Institute in Moscow.

\end{multicols}


\begin{thebibliography}{10}
\bibitem{ABZ99}
{Detailed consequences of the excess oxygen location
\protect\cite{Mehta94,Tranquada97} for the NMR/NQR spectrum will be dealt
with in a forthcoming paper.}
\bibitem{CM98}
{In \protect\cite{CM98m} the $m=1/2$ line splitting is wrongly ascribed to a
magnetic effect; the decomposition of the $m=1/2$ line into two lines must be
made in accordance with the $m=3/2$ line.}
\end{thebibliography}
\end{document}